\documentclass[reprint,
aps,
]{revtex4-1}
\usepackage{amsfonts}
\usepackage{amsmath}
\usepackage{amssymb,amscd}
\usepackage{psfrag}
\usepackage{graphicx}
\usepackage{braket}
\usepackage[dvips]{epsfig}
\usepackage{epsfig}
\usepackage{subfigure}
\usepackage{color}
\usepackage{amssymb}
\usepackage{hyperref}
\usepackage{lineno}

\hypersetup{colorlinks=true, citecolor=blue, urlcolor=blue, linkcolor=black}
\providecommand{\U}[1]{\protect\rule{.1in}{.1in}}

\begin{document}

\title{Protection of Noise Squeezing in a Quantum Interferometer with
Optimal Resource Allocation}
\author{Wenfeng Huang$^{1}$}
\author{Xinyun Liang$^{1}$}
\author{Baiqiang Zhu$^{1}$}
\author{Yuhan Yan$^{1}$}
\author{Chun-Hua Yuan$^{1,4}$}
\email{chyuan@phy.ecnu.edu.cn}
\author{Weiping Zhang$^{2,3,4,5}$}
\email{wpz@sjtu.edu.cn}
\author{L.Q.Chen$^{1,4}$}
\email{lqchen@phy.ecnu.edu.cn}

\begin{abstract}
Interferometers are crucial for precision measurements, including 
gravitational waves, laser ranging, radar, and imaging. The phase sensitivity,
the core parameter, can be quantum-enhanced to break the standard quantum 
limit (SQL) using quantum states. However, quantum states are highly fragile
and quickly degrade with losses. We design and demonstrate a quantum 
interferometer utilizing a beam splitter with a variable splitting ratio to 
protect the quantum resource against environmental impacts. The optimal 
phase sensitivity can reach the quantum Cram\'{e}r-Rao bound of the system. 
This quantum interferometer can greatly reduce the quantum source 
requirements in quantum measurements. In theory, with a 66.6$\%$ loss rate, 
the sensitivity can break the SQL using only a 6.0 dB squeezed quantum 
resource with the current interferometer rather than a 24 dB squeezed
quantum  resource with a conventional squeezing-vacuum-injected Mach-Zehnder
interferometer. In experiments, when using a 2.0 dB squeezed vacuum state, 
the sensitivity enhancement remains at $\sim$1.6 dB via optimizing the first splitting ratio when the loss rate changes from 0$\%$ to 90$\%$, indicating that the quantum  resource is
excellently protected with the existence of losses in practical 
applications. This strategy could open a way to retain quantum advantages 
for quantum information processing and quantum precision measurement in 
lossy environments. 
\end{abstract}

\date{\today }

\preprint{APS/123-QED} 

\address{$^1$State Key Laboratory of Precision Spectroscopy, Quantum Institute for Light and Atoms,
		Department of Physics and Electronic Science, East China Normal University, Shanghai 200062, China}
\address{$^2$School of Physics and Astronomy, and Tsung-Dao Lee Institute,
		Shanghai Jiao Tong University, Shanghai 200240, China}  
\address{$^3$Shanghai
		Research Center for Quantum Sciences, Shanghai 201315, P.R.China}  
\address{$^4$Shanghai Branch, Hefei National Laboratory, Shanghai 201315, China} 
\address{$^5$Collaborative Innovation Center of Extreme Optics, Shanxi
		University, Taiyuan, Shanxi 030006, China} 
\maketitle

Interferometers play a key role in precision measurements for gravitational
waves \cite{1-1,1-2,1-3,1-4}, gravity fields \cite{2-1,2-2,2-3,2-4}, imaging 
\cite{3-1,3-2,3-3}, and so on. Quantum squeezing states, including two-mode
squeezing \cite{4-1,4-2} and vacuum squeezing states \cite{4-3,4-4} are
important, which differentiate these measurements \cite%
{5-4,5-5,5-6,5-12,5-13} from classical measurements.
There have been numerous proof-of-principle experiments on noise-squeezing quantum
measurements in imaging \cite{5-7,5-8}, magnetometry \cite{5-9,5-10,5-11},
quantum information \cite{5-14,5-15}, etc. The typical example is the
utilization of squeezing vacuum as the input of Mach-Zehnder
interferometer(MZI) to improve the measurement sensitivity of gravitational
waves \cite{5-1,5-2,5-3}. However, in realistic situations, the measurement
system is rarely isolated and is inevitably affected by the environment \cite%
{6-1,6-2,6-3}. The losses are one of the main obstacles to the realization
of quantum measurement \cite{7-1,7-2}. The losses of quantum states bring
excess noise, resulting in quick degradation of the sensitivity. 

A typical quantum MZ interferometer is constructed of two input fields (a 
coherent field ${\varepsilon }_{0}$ and a squeezed vacuum ${S}_{0}$) and two
50$:$50 beam splitters for wave splitting and wave  recombination as shown in
Fig. 1(a), called
\begin{figure}[]
	\begin{centering}
		\includegraphics[width=0.9\columnwidth]{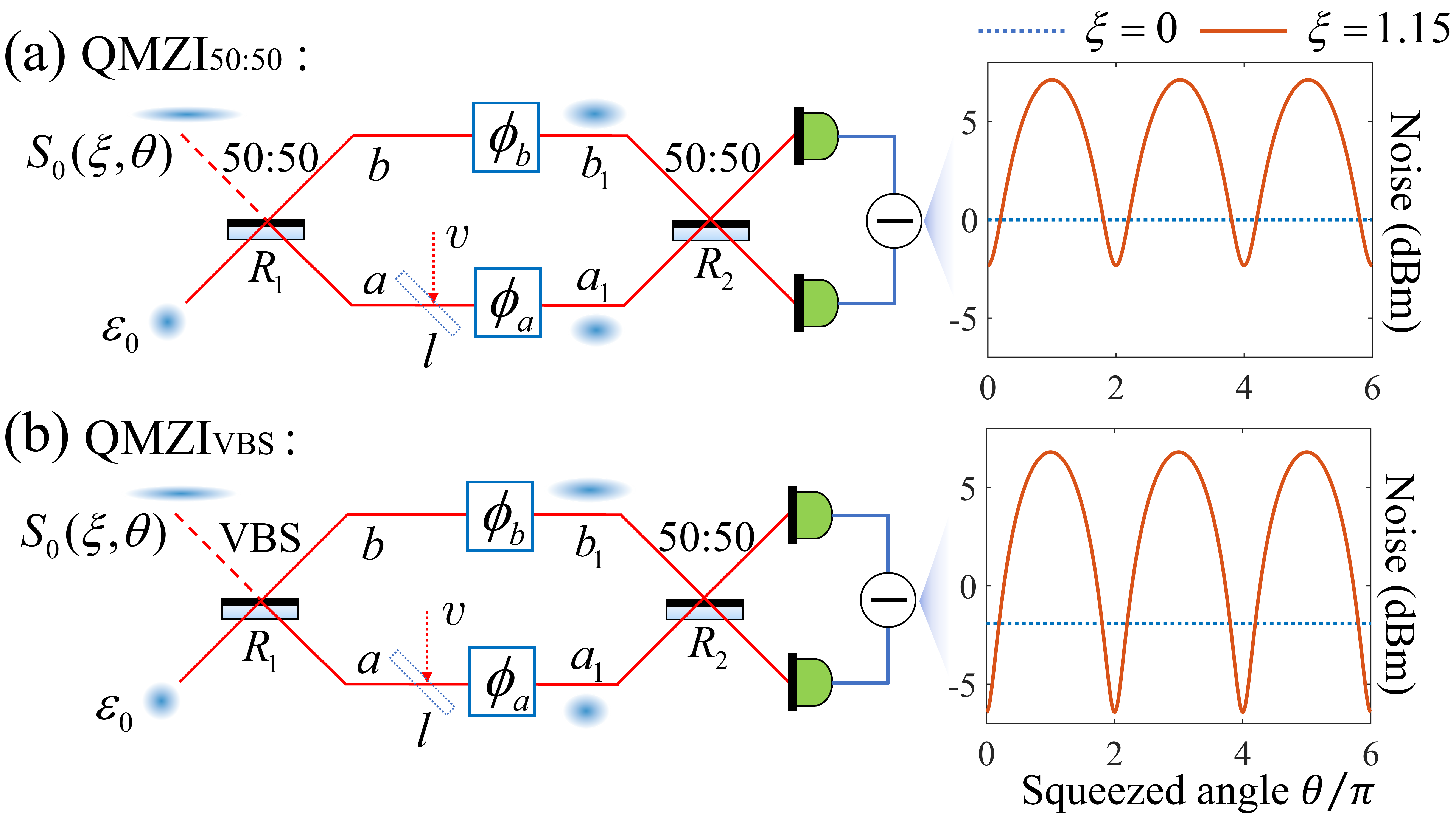}
		\par\end{centering}
	\caption{Schematic of lossy quantum Mach-Zehnder interferometer. $%
		\protect\varepsilon _{0}$: coherent field; $S_{0}$($\protect\xi ,\protect%
		\theta $): squeezed vacuum field with squeezing degree $\protect\xi $ and
		squeezing angle $\protect\theta $; $v$: vacuum field. $l$: internal loss
		rate of arm $a$; $\protect\phi _{a},\protect\phi _{b}$: phase shifts; $R_{1}$%
		, $R_{2}$: reflectivities of beam splitters. (a) QMZI$_{50:50}$ with
		two 50:50 beam splitters. (b) QMZI$_{\text{VBS}}$ with a variable
		beam splitter (VBS) and a 50$:$50 beam splitter. Blue shades: noise at the
		respective position of the interferometer. Two right insets show the noise
		squeezing of QMZI$_{50:50}$ and QMZI$_{\text{VBS}}$ with parameters $\protect\phi %
		_{a}-\protect\phi _{b}=\protect\pi /2$, $l=0.7$ and $\protect\xi =1.15$. }
\end{figure}
\begin{figure*}[tph]
	\begin{centering}
		\includegraphics[width=1.9\columnwidth]{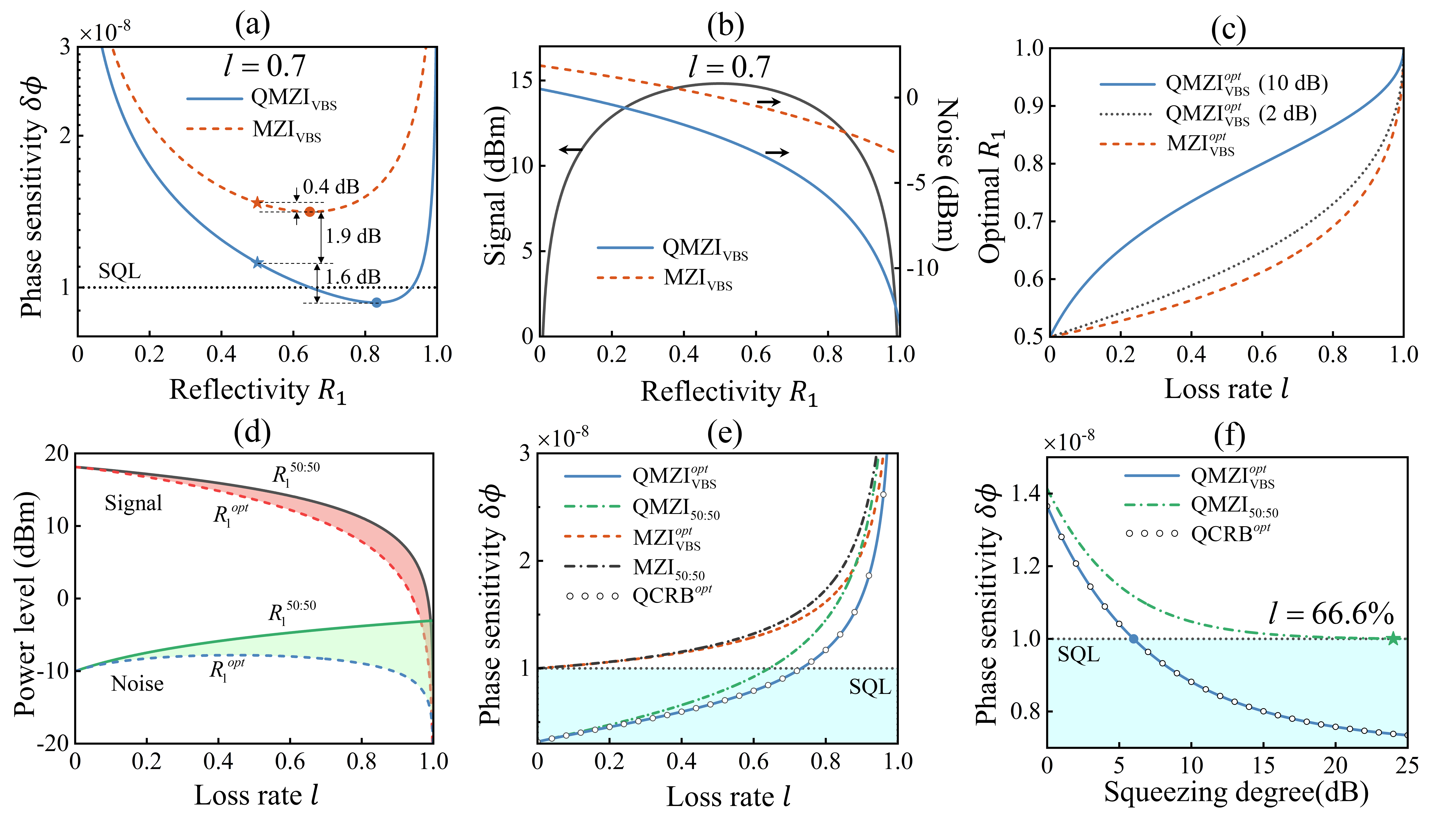}
		\par\end{centering}
	\caption{Theoretical analysis. (a) Sensitivity as a
		function of the reflectivity $R_{1}$. Solid and dashed lines represent QMZI$%
		_{\text{VBS}}$ and MZI$_{\text{VBS}}$. SQL: dotted line, lossless MZI$_{50:50}.$
		(b) Signal (left axis, solid line) and noise (right axis, blue
		solid line) of QMZI$_{\text{VBS}}$. Noise of MZI$_{\text{VBS}}$ (right axis, brown dashed
		line) is set to 0 dBm as a noise reference when $R_{1}=0.5$. (c) Optimal reflectivity $R_{1}$ for minimum
		sensitivity versus loss $l$ in MZI and QMZI (10 dB or 2 dB
		squeezing input). (d) Signal and noise as a function of loss rate $l
		$. QMZI$_{50:50}$: solid line; QMZI$_{\text{VBS}}$ : dashed line. Sensitivity as a
		function of (e) the loss rate $l$ and (f) the squeezing
		degree at $l=66.6\%$. QCRB$^{\text{opt}}$: quantum Cram\'{e}r-Rao bound with
		optimal $R_{1}$. QMZI$_{\text{VBS}}^{\text{opt}}$: QMZI$_{\text{VBS}}$ with optimal $R_{1}$. The SQL is obtained by $\frac{1}{\protect\sqrt{N}}$, with the total number of photons $N$ =$10^{16}$. In (a),(f),
		stars represent the sensitivity with $R_{1}=0.5$, and dots represent the sensitivity with optimal $R_{1}$. The squeezing degree of QMZI in (a),(b),(d),(e) is 10 dB.}
\end{figure*}
 QMZI$_{50:50}$. A 20\%  loss in one arm of this
interferometer results in a 3.5 dB decrease in the  sensitivity when using a
10 dB-squeezed vacuum, but results in only a 0.5 dB  decrease when using a
vacuum state instead ${S}_{0}$, that is a conventional  MZ interferometer
(MZI$_{50:50}$). The quantum enhancement of QMZI$_{50:50}$  is very fragile
with the existence of losses in the interference arm because  the advantage
of noise squeezing is eliminated with considerable losses \cite{8-1,8-2,8-3}.
Therefore, studying how to protect quantum resources in practical 
environments is fundamentally important.

In this Letter, to resist the loss effect, we propose and experimentally 
demonstrate a loss-tolerant quantum interferometer,\ called QMZI$_{\text{VBS}}$, as
shown in Fig. 1(b). The difference compared with QMZI$_{50:50}$ is a 
variable beam splitter (VBS) to realize wave splitting. The central purpose 
is to protect the quantum resource as much as possible to minimize the noise
and retain more of the coherent resource to maximize the signal and 
significantly improve the phase sensitivity in a lossy environment. A way to
realize this goal is to optimally allocate the quantum and coherent 
resources by adjusting the ratio $R_{1}$ of VBS.

The interference output is sensitive to the variation of the phase
difference between two arms when the phase difference is fixed at the
optimum point $\pi /2$, and a small phase shift $\Delta \phi $ ($\phi
_{a}-\phi _{b}=\pi /2+\Delta \phi $) is applied. Then, the phase sensitivity
of QMZI$_{\text{VBS}}$ is given as (Supplementary Material, Sec. II \cite{S})  
\begin{equation}
\delta \phi =\sqrt{\frac{(1-R_{1})l+(1-l)e^{-2\xi }}{4(1-R_{1})R_{1}(1-l)N}},
\end{equation}%
which depends on the loss rate $l$, reflectivity $R_{1}$, squeezing degree $%
\xi $ and photon number $N$ in the input laser $\varepsilon _{0}$.
Obviously, a large squeezing degree and a small loss are both beneficial for
sensitivity. Therefore, reducing the loss by improving the reflectivity of
the optical mirror and the vacuum in the optical path is an effective method 
\cite{10-1}, as is utilizing better quantum resources, which is an area that
many groups have made great efforts in \cite{11-1,11-2,12-1}. 
However, isolation of the surrounding environment and improvement of quantum
resources are difficult to achieve in experiments. Furthermore, the quantum
enhancement of the sensitivity for QMZI$_{50:50}$ guaranteed by the quantum
resource quickly degrades with the loss. For a given $\xi $, there is a special loss rate that reduces the sensitivity of QMZI$_{50:50}$ to the standard quantum limit (SQL, defined as the
sensitivity of lossless MZI$_{50:50}$), called the loss rate limit $l_{\text{SQL}}=2(e^{2\mathbf{\xi }}-1)/(3e^{2\mathbf{\xi }}-2)$ \cite{S}. With $\xi =$ 0.69 (6 dB squeezing) or 2.76 (24 dB
squeezing), the corresponding $l_{\text{SQL}}$ is 0.6 or 0.666, respectively,
indicating that improvement of the quantum resource contributes little to
the enhancement of sensitivity in a lossy environment. How to protect the
quantum advantage to achieve the best sensitivity with the use of limited
resources is the focus of attention in quantum information processing and
quantum measurement.

In addition to the loss rate and squeezing degree, there is a parameter $%
R_{1}$ in Eq. (1), which can be adjusted to protect the quantum
resource when $\xi \neq 0$ and reallocate the coherent resource
in two arms when $\xi =0$. As shown in Fig. 2(a), when $l=0.7$, the sensitivity of QMZI$_{50:50}$ (star on solid line) is above the SQL by 0.9 dB and the quantum enhancement is 2.3 dB compared with MZI$%
_{50:50}$ (star on dashed line). By optimizing $R_{1}$, the sensitivity of MZI$_{\text{VBS}}$ is slightly improved by 0.4 dB compared to MZI$_{50:50}$.
However, the sensitivity of QMZI$_{\text{VBS}}$ is significantly improved by 1.6 dB compared with QMZI$_{50:50}$. Furthermore, QMZI$_{\text{VBS}}$ is improved to below SQL by 0.7 dB, which is better than the best MZI$_{\text{VBS}}$ result by 3.5 dB. Figure. 2(a) clearly indicates that $R_{1}$ optimization retains the quantum advantage of noise squeezing well.

Such significant quantum enhancement simply by adjusting $R_{1}$ is due to
the optimal balance between signal and noise in a lossy environment. The
ratio $R_{1}$ determines the proportion of the quantum resource and coherent
field in the lossless path and lossy path. A larger $R_{1}$ tends to pass
more of the quantum resource into the lossless path to maintain the noise
squeezing and simultaneously more of the coherent field into the lossy path,
resulting in a decrease in the effective phase-sensitive photon number. At a
certain loss $l$, the signal and the noise are 
\begin{eqnarray}
\text{signal} &=&2\sqrt{(1-R_{1})R_{1}(1-l)}N\Delta \phi , \\
\text{noise} &=&\sqrt{[(1-R_{1})l+(1-l)e^{-2\xi }]N},
\end{eqnarray}%
which exhibit different dependence on $R_{1}$ \cite{S}. As shown in Fig. 2(b), the maximum signal always appears at $%
R_{1}=0.5$, but the noises of both QMZI$_{\text{VBS}}$ and MZI$%
_{\text{VBS}}$ monotonically decrease with $R_{1}$. Furthermore, the noise initially decreases faster, and then, the signal is
reduced more quickly. In general, the noise of QMZI$_{\text{VBS}}$ decreases
faster than that of MZI$_{\text{VBS}}$ due to the protection of quantum resource. The optimal $R_{1}$ for the minimum $\delta \phi $
can be given as
\begin{equation}
R_{1}^{\text{opt}}=\left\{ 
\begin{matrix}
\frac{B-\sqrt{B^{2}-Bl}}{l} & (l\neq 0,B=(1-l)e^{-2\xi }+l) &  \\ 
0.5 & (l=0) & .%
\end{matrix}%
\right. 
\end{equation}%
In Fig. 2(c), $R_{1}^{\text{opt}}$ grows with $l$, whose growth rate
depends on the squeezing degree $\xi $. MZI$_{\text{VBS}}^{\text{opt}}$ curve
corresponds $\xi =0$. Better squeezing degree $\xi $ requires a larger $%
R_{1}^{\text{opt}}$ because quantum resources with higher performance are more
fragile in a lossy environment. This is the reason why quantum measurement
is difficult to apply in practice.

In general, the loss has a negative effect on the sensitivity due to the
decrease of phase-sensitive photon number and the reduction of quantum
performance. However, with the optimal $R_{1}$, both the sensitivity and
quantum enhancement can be significantly improved. In Fig. 2(d), when the
loss increases, the signal reduces while the noise increases for QMZI$%
_{50:50}$, but both the signal and noise decrease for the optimal QMZI$_{\text{VBS}}$,
which differentiates QMZI$_{\text{VBS}}$ from QMZI$_{50:50}$. Therefore, the sensitivity of QMZI$_{\text{VBS}}^{\text{opt}}$ is always better than QMZI$_{50:50}$ and MZI$_{\text{VBS}}^{\text{opt}}$ [Fig. 2(e)], clearly
revealing the advantage of an adjustable $R_{1}$ in a lossy
environment. Here, we show the results of sensitivity improvement via the $R_{1}$
optimization when the loss of arm $b$, $l_{b}=0$. In fact, in
practical applications, both arms may suffer losses simultaneously.
However, sensitivity improvement can still be achieved via $R_{1}$
optimization with the same trend as the results of $l_{b}=0$ \cite{S}. Furthermore, in many applications, arm $b$ can be
retained in a local environment that can be controlled to be nearly
lossless, so we theoretically study and experimentally demonstrate the
optimization effect in a simple and clear way.  
\begin{figure}[tb]
\includegraphics[width=1\columnwidth]{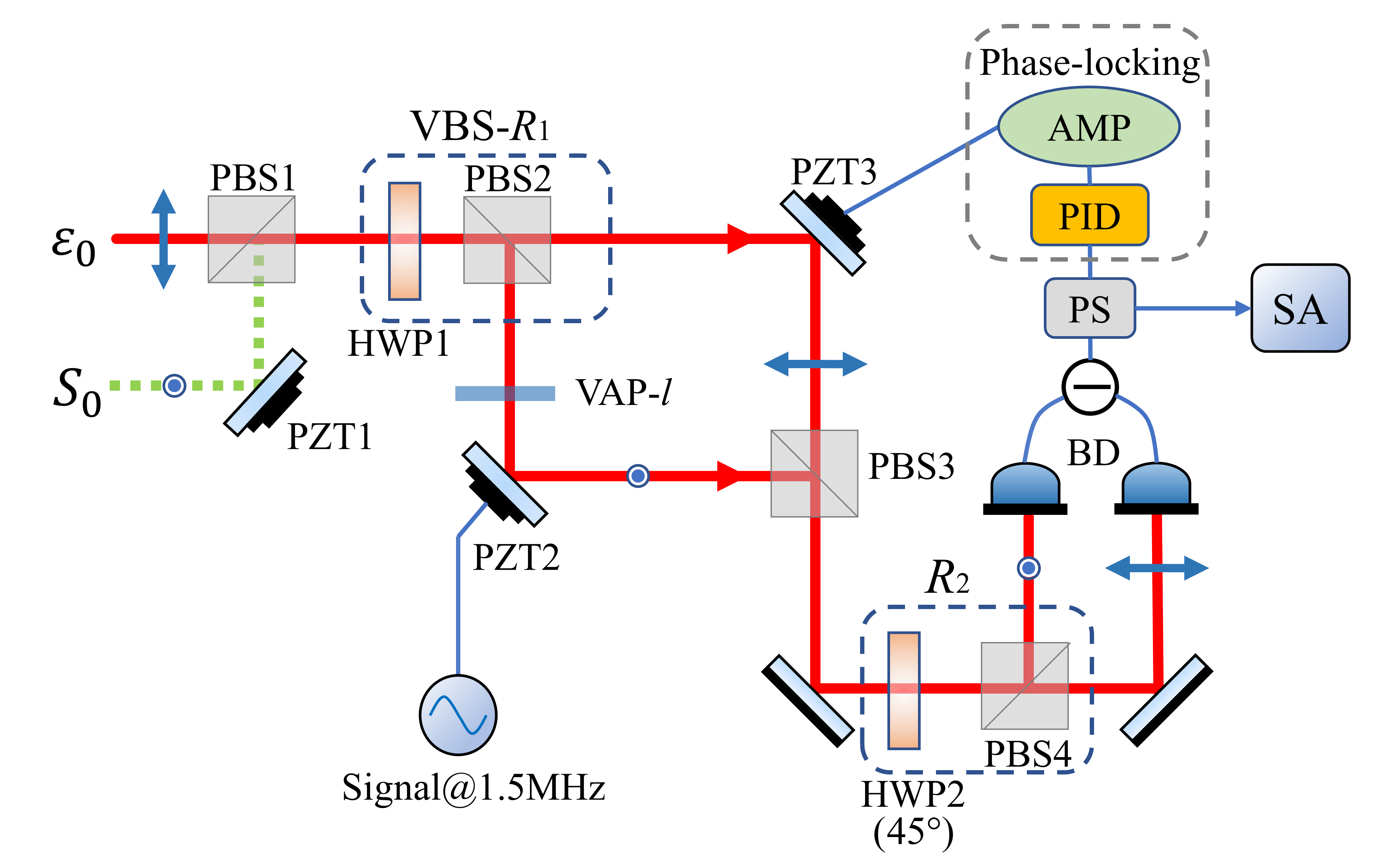}  
\caption{Experimental setup. Coherent field ${\protect\varepsilon }%
_{0}$ with horizontal polarization and squeezed vacuum field ${S}_{0}$ with
vertical polarization. The power of ${\protect\varepsilon }_{0}$ is 2.5 mW. The squeezing degree of ${S}_{0}$ is 2 dB. The interference signal is measured by balanced
intensity-difference detection (BD) and analyzed by a spectrum analyzer
(SA). Polarization beam splitter (PBS); variable attenuation plate (VAP); half-wave plate (HWP); variable beam splitter (VBS); amplifier (AMP); power splitter (PS); piezoelectric transducer (PZT). }
\end{figure}
\begin{figure*}[tph]
\includegraphics[width=1.8\columnwidth]{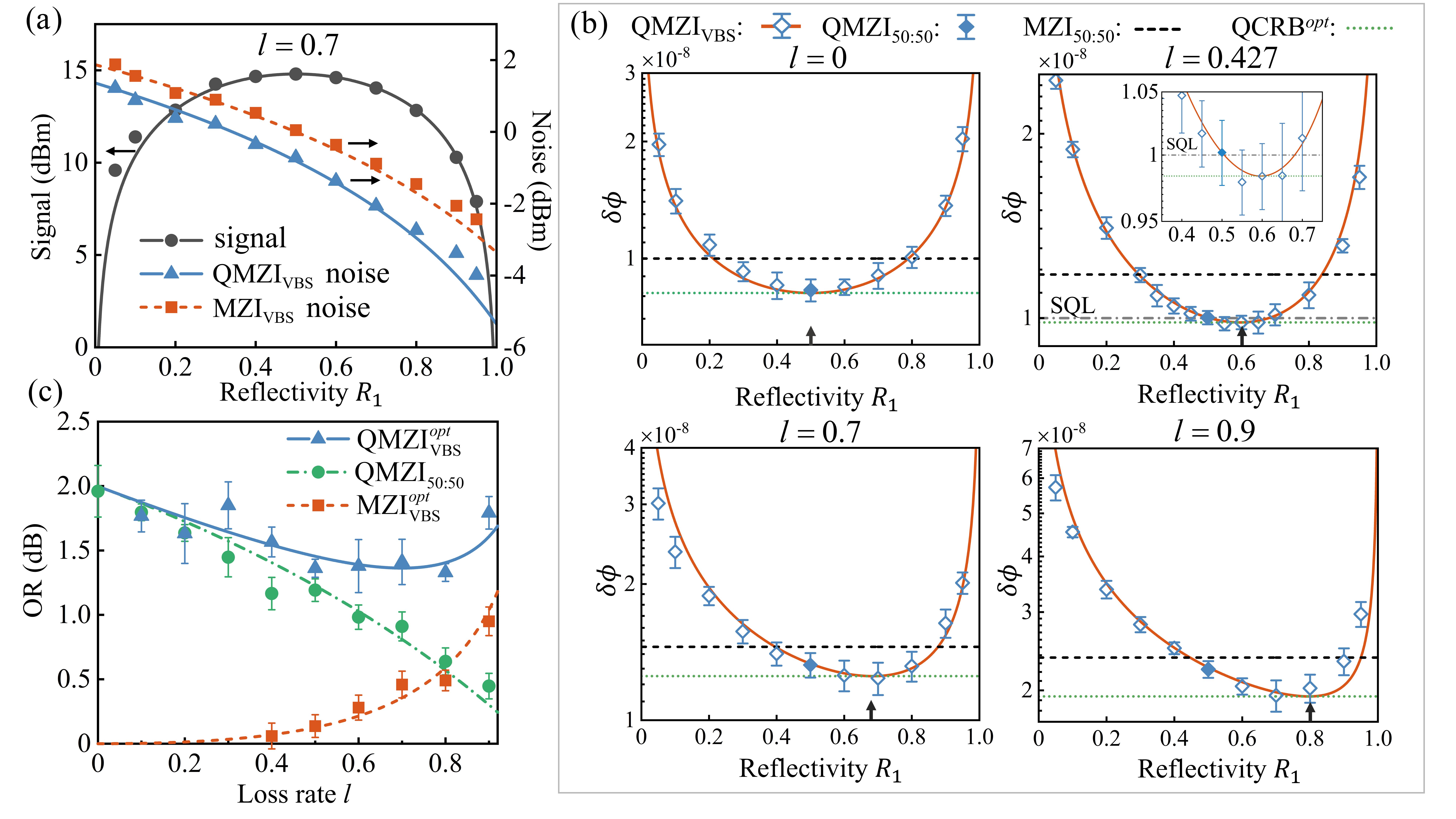}  
\caption{Experimental results. Squeezing degree of ${S}_{0}$ is 2.0 dB. (a) Signal and noise versus reflectivity $R_{1}$
at $l=0.7$. Blue triangles and black dots are the noise and signal of QMZI$%
_{\text{VBS}}$, respectively. Brown squares: noise of MZI$_{\text{VBS}}$. Lines represent
theoretical fittings. (b) Sensitivity versus reflectivity 
$R_{1}$ with different losses. The arrows
indicate the location of optimal $R_{1}$. (c) Optimization ratio (OR) is defined as -20log$_{10}(\protect\delta \protect\phi _{i}/\protect\delta \protect\phi_{\text{MZI}_{50:50}})$, where $i$ is $\text{QMZI}_{\text{VBS}}^{\text{opt}}$, $\text{QMZI}_{50:50}$, or $\text{MZI}_{\text{VBS}}^{\text{opt}}$.}
\end{figure*}

Another advantage of the current interferometer is that the performance 
requirements for quantum sources can be greatly reduced. In the QMZI$
_{50:50} $ case, the improvement of the quantum resource helps little in 
enhancing the phase sensitivity $\delta \phi $ in a  lossy environment.
According to the loss rate limit $l_{\text{SQL}}$, when the loss  rate $l\geq 2/3$, 
$\delta \phi $ is worse than the SQL regardless of the  squeezing degree of
the squeezing vacuum. However, in the QMZI$_{\text{VBS}}$ case,  $\delta \phi $ can
be better than the SQL by optimizing $R_{1}$ at $ l\geq 2/3$ using a
squeezing vacuum with a relatively poor squeezing degree.  In Fig.
2(f), when $l$ is 66.6$\%$, the achievement of  SQL-broken
sensitivity requires 24 dB squeezing with 50$:$50 beamsplitters,  while only
6.0 dB squeezing is required with the optimal VBS of $R_{1}$=0.72.

To note that, in Figs. 2 (e,f), the quantum Cram\'{e}r-Rao bound (QCRB) 
curve is also given to evaluate the performance of interferometers
\cite{S}. QCRB is the detection-independent
ultimate limit of  the sensitivity \cite{15-1,15-2,15-3}. The optimal
sensitivity of QMZI$_{\text{VBS}}$  can saturate the optimal QCRB at all loss rates
and squeezing degrees, showing that the  sensitivity of QMZI$_{\text{VBS}}$ reaches
the ultimate limitation.

In the experiment, the VBS of QMZI$_{\text{VBS}}$ consists of a polarization 
beam splitter (PBS2) and a half-wave plate (HWP1), as shown in Fig. 3. $R_{1}$ can be changed by adjusting HWP1. ${S}_{0}$ and ${\varepsilon }_{0}$,
spatially overlapping with orthogonal polarizations, are divided into  two
interference arms after passing through VBS-$R_{1}$ and recombined to 
realize a quantum interferometer. HWP2 and PBS4 are utilized as the 
traditional 50:50 beam splitter. The visibility of 
lossless MZI$_{50:50}$ is 99$\%$. A variable attenuation plate is placed in one interference arm to change the loss rate $l$. The phase 
difference between two arms is locked at $\pi /2$. Squeezing degree of $
S_{0} $ is 2.0 dB, generated via polarization self-rotation 
effect in atomic vapor \cite{13-1,13-2}.

The dependences of the signal, noise and sensitivity on $R_{1}$ were
measured to further experimentally explore the physical mechanism for $R_{1}$
optimization. In Fig. 4(a), with the increase of $R_{1}$, the noise
level of QMZI$_{\text{VBS}}$ and that of MZI$_{\text{VBS}}$
monotonically decrease and the best signal appears at $R_{1}=0.5
$, matching the theoretical results well. In particular, as $%
R_{1}$ increases, the noise of QMZI$_{\text{VBS}}$ drops faster than that of MZI$%
_{\text{VBS}}$ due to the protection of quantum noise. The four figures in Fig.
4(b) show the sensitivity as a function of $R_{1}$ at different losses. The optimal $R_{1}$ is 0.5, 0.6, 0.66, and 0.8 for loss rates $l=$0, 0.427, 0.7, and 0.9, respectively.
A larger loss requires a larger optimal $R_{1}$. In
general, the absolute sensitivities of both MZI$_{50:50}$ and QMZI$_{\text{VBS}}$
increase with the loss, mainly due to the decrease in the number of
phase-sensitive photons. The best sensitivity of QMZI$_{\text{VBS}}$ is always
better than that of QMZI$_{50:50}$ and reaches the optimal QCRB
because of the protection of the quantum resource via $R_{1}$ optimization.
In particular, at $l=0.427$, the phase sensitivity of QMZI$_{50:50}$ is
worse than SQL. By adjusting R$_{1}$ to the optimal value, the sensitivity
of QMZI$_{\text{VBS}}^{\text{opt}}$ can still go beyond SQL.

The optimization ratios (ORs) of QMZI$_{\text{VBS}}^{\text{opt}}$/MZI$_{50:50}$ and MZI$_{\text{VBS}}^{\text{opt}}$/MZI$_{50:50}$ at different losses are given in Fig. 4(c) to show the optimization degree. The results of QMZI$_{50:50}$ are also compared under the same initial conditions. The sensitivities of the two
quantum interferometers (QMZI$_{50:50}$ and QMZI$_{\text{VBS}}^{\text{opt}}$) are both beyond
the SQL by 2.0 dB at $l=0$. With the growth of the loss from 0 to
90\%, the OR monotonically decreases to 0.5 dB in the
QMZI$_{50:50}$ case, while it is maintained at 1.6 dB in the QMZI$_{\text{VBS}}^{\text{opt}}$
case. It experimentally demonstrates that the conventional quantum
interferometer is very sensitive to losses, but the utilization of an
adjustable beam splitter to adjust $R_{1}$ is an effective way to optimize the
sensitivity of quantum measurements in a lossy environment. As a result, QMZI$_{\text{VBS}}^{\text{opt}}$ is loss tolerant. Furthermore, we emphasize that in physics varying the beam splitter ratio optimizes the performance of the interferometer in the presence of losses for both vacuum and squeezing-vacuum injections, as shown in Fig. 4(c). While in the case of squeezing-vacuum injection, squeezing protection through the allocation in two arms of the QMZI$_{\text{VBS}}^{\text{opt}}$ shows better performance compared with vacuum injection case \cite{S}.

So far, we demonstrate the quantum protection of QMZI$_{\text{VBS}}$ with the
loss in one arm. In fact, such a proposal can also be applied to improve the
performance with the existence of internal losses in two arms.  As
long as the losses are unbalanced, whether with or without
detection loss \cite{14-1,14-2}, both the sensitivity and quantum protection
can be improved by the $R_{1}$ optimization.  Current QMZI$_{\text{VBS}}^{\text{opt}}$ is better
than QMZI$_{50:50}$ \cite{S}. We
have proven the  effectiveness of the QMZI$_{\text{VBS}}$ model.

The purpose of introducing the VBS into a quantum interferometer is to 
protect the quantum resource as much as possible to significantly improve 
the phase sensitivity in a lossy environment. The phase sensitivity at the optimal $R_{1}$  reaches the ultimate limit, QCRB$^{\text{opt}}$, and is much better than that with QMZI$_{50:50}$. The performance  requirements for quantum sources are greatly reduced, alleviating the  vulnerability of the quantum resource to some extent. In experiments, when  using a 2.0 dB-squeezed vacuum, as high as 1.6 dB sensitivity enhancement is retained at the loss rate from 0 to 90$\%$. 
In practical applications, quantum  resources are difficult to improve, but
coherent resources are much easier  to achieve. The current technique
promises full use of quantum resources in  a practical environment, which
has important application prospects in  quantum precision measurement,
quantum information, and biotechnology.

This work is supported by the Innovation Program for Quantum Science and Technology 2021ZD0303200; the National Natural Science Foundation of China Grants No. 12274132, No. 11974111, No. 12234014, No. 11654005, No. 11874152, and No. 91536114; Shanghai Municipal Science and Technology Major Project under Grant No. 2019SHZDZX01; Innovation Program of Shanghai Municipal Education Commission No. 202101070008E00099; the National Key Research and Development Program of China under Grant No. 2016YFA0302001; and Fundamental Research Funds for the Central Universities. W.Z. acknowledges additional support from the Shanghai Talent Program. 

\providecommand{\noopsort}[1]{}\providecommand{\singleletter}[1]{#1}%

\end{document}